\documentclass[prl,aps,twocolumn]{revtex4}
\usepackage{graphicx}
\begin{document}
\title{Transmission of phase information between electrons and holes in graphene}
\author{Atikur Rahman}
\author{Janice Wynn Guikema}
\author{Soo Hyung Lee}
\author{Nina Markovi\'c}
\affiliation{{\it Department of Physics and Astronomy, Johns Hopkins
University, Baltimore, Maryland 21218, USA.}}
\begin{abstract}
We have studied quantum interference between electrons and holes in a split-ring gold interferometer
with graphene arms, one of which contained a pn junction. The carrier type, the pn junction and
the phase of the oscillations in a magnetic
field were controlled by a top gate placed over one of the arms. We observe clear Aharonov-Bohm
oscillations at the Dirac point and away from it, regardless of the carrier type in each arm.
We also find clear oscillations when one arm of the interferometer
contains a single pn junction, allowing us to study the interplay of Aharonov-Bohm effect and Klein
tunneling.
\end{abstract}
\maketitle
The observation of Aharonov-Bohm (AB) oscillations \cite{ab_0} is a
straightforward way of demonstrating phase coherence in mesoscopic
samples. When an external out-of-plane magnetic field is applied to
a mesoscopic ring, the phase coherent charge carrier trajectories
encircling the ring acquire a phase difference $\Delta \phi = 2\pi
eBS/h$ (where $e$ is the electron charge, $h$ is Planck's constant,
and $S$ is the area of the ring) due to the presence of the magnetic
field $B$. Consequently, the low temperature magnetoresistance shows
an oscillation with a characteristic period $\Delta B = \Phi_0/S$,
where $\Phi_0=h/e$ is the flux quantum. Such oscillations are
commonly found in mesoscopic rings of metals \cite{ring_1},
semiconducting heterostructures \cite{ring_3}, carbon nanotubes
\cite{cnt_1,cnt_2} and topological insulators \cite{nanoribbon}.
Electron transport through graphene is peculiar in many ways
\cite{gr_1, gr_2, gr_3,eh_1}, particularly at the Dirac point where
pseudodiffusive transport is expected in clean samples
\cite{Katsnelson_1}, and electron-hole puddles have been observed in
scanning probe measurements \cite{puddle_5}. The AB effect has been
studied in mesoscopic graphene rings, but no oscillations were found
at the Dirac point (charge neutrality point where the density of
states is zero and the conductance is minimal) \cite{graphene_1,
graphene_2, graphene_3, graphene_4}. The failure to observe the AB
oscillations at the Dirac point is most likely due to the decrease
of the phase coherence length in that regime, which is exacerbated
by the inevitable presence of edges in the rings, which can act as a
source of disorder \cite{edge_1, edge_2, edge_3}. In this work, we
used a split-ring interferometer \cite{linear_1} to investigate
phase coherence in graphene. In this type of interferometer, the
gold leads reach deep into the graphene ring, such that each arm
contains only a short graphene section.  AB oscillations in a
split-ring geometry have been used before to determine the phase
shift of an electron traversing a quantum dot \cite{ab_2}, and to
study the magneto-electric AB effect \cite{ab_1, ab_3}.
Topologically equivalent to a graphene ring, the split-ring gold
interferometer with graphene arms allows us to study the quantum
coherence in graphene in the quasi-ballistic regime, demonstrate
phase coherence between electrons and holes and observe
phase-coherent tunneling through a pn junction.

\begin{figure}
  \includegraphics[width=7cm]{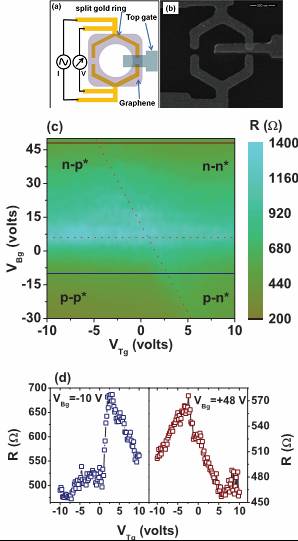}
  \caption{(a) Device schematic and the measurement configuration.
  (b) Scanning electron microscope image of a device with a $\sim$ 85 nm wide graphene
  region completing the split gold ring. (c) Resistance as a function of back
  gate voltage and top gate voltage. All four regions (n-p$^*$, n-n$^*$, p-n$^*$
  and p-p$^*$) are clearly visible. Data corresponding to $-$10 V and +48 V back gate
  voltage (blue and red lines) are plotted separately in (d).}
\end{figure}

The device geometry is shown in Fig. 1(a) and (b). The split gold ring
was deposited on top of a single layer graphene flake. A hole was etched in graphene
in the center of the device, making sure that the diameter of the
removed portion is such that the graphene edge is not very close to the edges
of the gold leads on each side. A top gate (Au) was deposited on top of an insulating
layer (Al$_2$O$_3$) over one arm of the ring (see supporting
material for fabrication details).
With the help of the back gate, which affects the entire
device, and the top gate, which affects only one arm, we can
manipulate the carrier type and concentration in the two arms of the
ring independently. In this measurement configuration, the contact
resistance at the graphene-gold interface cannot be bypassed and the
measurements were done in a quasi four-probe geometry. However, the
observed low resistance ($< 300\; \Omega$) suggests low contact
resistance in these devices. As the resistance of the graphene is
much larger than the resistance of the metallic portions, the
voltage drop occurs mainly across the graphene region. A small
change in graphene resistance dominates the overall response and is
easily detected.

A scanning electron microscope image of a device with a $\sim$ 85 nm
long graphene sections in each arm and a top gate on one arm is shown in
Fig. 1(b). Measurements
were done in a $^3$He cryostat equipped with an 8 Tesla magnet at
250 mK base temperature. Electrical measurements were done in a
current-biased configuration by driving a 10 nA current of 17 Hz
through the current leads, and the generated voltage was measured in
a quasi 4-probe geometry (as indicated in Fig. 1(a)) using an analog
lock-in amplifier. In Fig. 1(c) we show the resistance of a typical
device as a function of top gate voltage ($V_{Tg}$) for various back
gate ($V_{Bg}$) voltages. With zero top gate voltage, the Dirac
point was located at $V_{Bg}=6$ V. It is clear from the data that,
by suitable choice of back gate and top gate voltages, the graphene
portions in the two arms of the ring can be made both
electron-type, both hole-type, or mixed (i.e, the charge carriers in one
arm are electrons, and the charge carriers in the other arm are
holes). We will mark the types of charge carriers in the two arms
p-p$^*$, p-n$^*$, n-p$^*$ or n-n$^*$ (where p and n denote the
carrier type in the arm with no top gate, while p$^*$ or n$^*$
denote the carrier type in the top-gated arm with non-zero
$V_{Tg}$).

The resistance of the device as a function of top gate voltage is
shown in Fig. 1(d) for two fixed values of back gate voltage ($-$10
V and +48 V).

\begin{figure}
  \includegraphics[width=7cm]{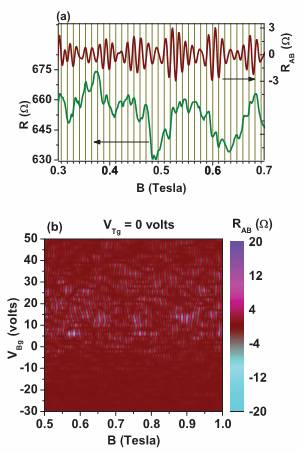}
\includegraphics[width=7cm]{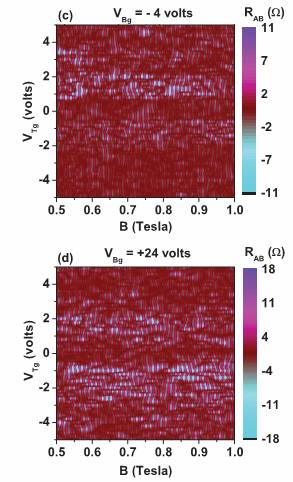}

  \caption{(a) Resistance and the oscillating part of the resistance
  as a function of magnetic field for $V_{Bg}$ = $-$4 V and zero $V_{Tg}$.
  The average periodicity (marked by vertical lines) is $\approx 12$ mT.
  $R_{AB}$ as a function of magnetic field and (b) back gate
  voltage for $V_{Tg}$ = 0 V, and top gate voltage for (c) $V_{Bg}$ = $-$4 V
  and (d) $V_{Bg}$ = +24 V. Clear oscillations are visible for
  all combinations of back gate and top gate voltages.}
\end{figure}

To investigate the Aharonov-Bohm oscillations, we scanned the
magnetic field at a ramp rate of 5 mT/min. The magnetoresistance at
250 mK shows a pronounced periodic oscillation on top of a slowly
varying aperiodic background. The aperiodic background arises due to
conductance fluctuations originating from the penetration of the
magnetic field in the arms \cite{apdc}. Using Fourier analysis, we
filtered out the aperiodic conductance fluctuations and obtained the
Aharonov-Bohm flux-dependent part of the resistance ($R_{AB}$). Fig.
2(a) shows magnetoresistance data for $V_{Bg}$ = $-$4 V with zero
top gate voltage (the graphene sections in both arms of the
interferometer are p-type in this case). A pronounced resistance
oscillation is visible in the raw data. The oscillating part of the
resistance shows that the period of $R_{AB}$ is about 12 mT. For
$h/e$ Aharonov-Bohm oscillation, the observed period corresponds to
a ring with a radius of 330 nm which is in reasonable agreement with
the radius of our device. At a bias current of 10 nA, the
peak-to-peak amplitude of $R_{AB}$ is up to $\sim 3$\% of the total
resistance.

In Fig. 2(b) we show the variation of $R_{AB}$ as a function of
magnetic field for various back gate voltages with zero top gate
voltage. It is evident that oscillations are observed over the
entire range of back gate voltage and magnetic field (also in
supporting material, Fig. S3).  For a fixed magnetic field, $R_{AB}$
shows oscillatory behavior as a function of gate voltage. In Fig.
2(c) we show $R_{AB}$ as a function of magnetic field for various
top gate voltages and a fixed ($-$4 V) back gate voltage. It is
important to note that application of the top gate voltage does not
destroy the Aharonov-Bohm oscillations: they are clearly visible
even when the carrier types in the two arms of the ring are
different. Observing the Aharonov-Bohm oscillations clearly
demonstrates quantum interference between electron-type states in
graphene in one arm and the hole-type states in the other arm.
Similar results were obtained for a fixed positive back gate voltage
as a function of top gate voltage (Fig. 2(d)). We find pronounced
Aharonov-Bohm oscillations for all possible combinations of the
carrier types in the two arms: p-p$^*$, n-n$^*$, p-n$^*$ and
n-p$^*$.

We do not observe any higher harmonics (those with periodicity
$\Phi_0/n$ where $n$ is an integer) as a function of top gate
voltage for any value of the back gate voltage. Also, the nature of
the oscillations remains the same at relatively high magnetic field
(up to $8$ Tesla, the maximum field used in our measurements). This
indicates that the main contribution to the Aharonov-Bohm
oscillations comes from the direct paths through the arms of the
ring \cite{ab_2, ab_3}. The overall resistance increases with
increasing magnetic field and the oscillations survive even at high
field, but no significant reduction or increase of the oscillation
amplitude was observed as a function of the magnetic field
(supporting material, Fig. S4), in contrast to previous observations
\cite{graphene_1, graphene_2, graphene_3}. An increase in the
oscillation amplitude has been observed previously in graphene
nanorings in the high field region \cite{graphene_1}, and was
suggested to be due to orbital effects \cite{graphene_1, graphene_2}
or scattering on magnetic impurities \cite{graphene_2}.

\begin{figure}
  \includegraphics[width=8cm]{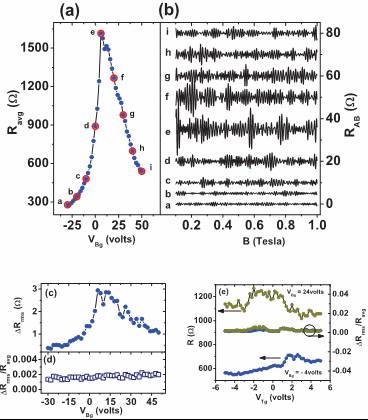}
  \caption{(a) Average resistance (averaged over the entire
  magnetic field scan from 0 to 1.05 T) as a function of back gate voltage
  for zero top gate voltage. The $R_{AB}$ values measured at different back gate voltages (indicated
  by solid circles in (a)) are shown in (b), offset for clarity. (c) $\Delta R_{rms}$ as a
  function of back gate. It is clear that it follows the same trend as $R_{avg}$ shown in (a).
  (d) $\Delta R_{rms}/R_{avg}$ as a function of $V_{Bg}$. (e) Variation of
  $R$ and $\Delta R_{rms}/R_{avg}$ as a function of $V_{Tg}$ for $V_{Bg}$ = 24 and $-$4 V.}
\end{figure}

In contrast to the previous work on graphene rings \cite{graphene_1,
graphene_2, graphene_3, graphene_4}, we find pronounced
Aharonov-Bohm oscillations at the Dirac point (shown in Fig. 3(a),
3(b) and supporting material, Fig. S3). The r.m.s. value of the
oscillation amplitude ($\Delta R_{rms}$) increases as the back gate
voltage approaches the Dirac point and becomes maximum at that point
(as shown in Fig. 3(b) and (c)). Fig. 3(d) shows $\Delta
R_{rms}/R_{avg}$, where $R_{avg}$ is the resistance averaged over
the entire magnetic field scan for fixed back and top gate voltages,
as a function of back gate voltage for zero top gate voltage. It is
clear from the data that the value of $\Delta R_{rms}/R_{avg}$ is
mostly independent of the back gate voltage. This indicates that the
visibility of $R_{AB}$ remains unaffected by the back gate voltage
and $\Delta R_{rms}$ shows a linear dependence on $R_{avg}$
(supporting material, Fig. S5) \cite{linear_1, graphene_1,
linear_2}. Similar results are obtained as a function of top gate
voltage for a fixed back gate voltage (Fig. 3(e)). For a fixed
(positive or negative) back gate voltage $\Delta R_{rms}/R_{avg}$
shows similar behavior, with similar amplitude for both positive and
negative top gate voltages. We conclude that the nature and the
visibility of the Aharonov-Bohm oscillation remain the same,
regardless of whether the carrier types are the same (n-n$^*$ or
p-p$^*$) or different (n-p$^*$ or p-n$^*$) in the two arms of the
interferometer.

It has been demonstrated that electron-hole puddles are present at the
charge neutrality point and that these puddles contribute to
the observed non-zero conductivity at the Dirac point
\cite{puddle_1, puddle_2, puddle_3, puddle_4, puddle_5}. The typical size of
the puddles is found to be on the order of 20nm, which is smaller than the gap
between the gold electrodes in our samples.
Our observation of pronounced Aharonov-Bohm oscillations at the Dirac
point shows that the charge transport through these puddles
preserves phase coherence. This is also consistent with our
observation of phase coherence away from the Dirac point, in the
case when the charge carriers are different in the two arms of the
interferometer.

Furthermore, the edge of the top gate on some samples (such as the one pictured in Fig. 1.b.)
was between the gold electrodes, allowing us to create a pn-junction
in one of the arms of the interferometer. Clear AB oscillations were observed in that case
as well. The amplitude of the oscillations was slightly asymmetric as a function of the top
gate voltage, which might be expected for Klein tunneling in smooth pn junctions \cite{Recher_1, Schelter}.

\begin{figure}
  \includegraphics[width=8cm]{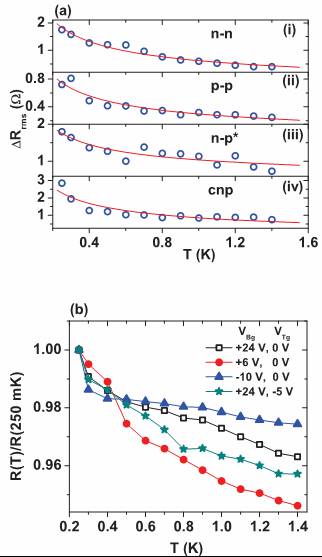}
  \caption{(a) Temperature dependence of the $\Delta R_{rms}$ for various back gate and top gate voltages.
  (i) $V_{Bg}$ = 24 V, $V_{Tg}$ = 0; both the arms have n-type carriers (n-n$^*$).
  (ii) $V_{Bg}$ = -10 V, $V_{Tg}$ = 0; both arms are p-type (p-p$^*$).
  (iii)$V_{Bg}$ = 24 V, $V_{Tg}$ = -5 V; one arm is n-type, and the other arm is p-type (n-p$^*$).
  (iv) $V_{Bg}$ = 6 V, $V_{Tg}$ = 0; charge neutrality point (cnp).
  (b) Variation of resistance (normalized by the 250 mK values) as a function of
  temperature has been shown for the above combinations of $V_{Bg}$ and $V_{Tg}$}
\end{figure}

The observation of phase coherence in the interferometer is further
supported by the measurements of the temperature dependence of the
oscillation amplitude for various back gate and top gate voltages.
We found that, irrespective of all possible combinations of the back
gate and top gate voltages, $R_{AB}$ shows a $T^{-1/2}$ dependence
(shown in Fig. 4(a)) between 250 mK and 1.4 K. Such temperature
dependence of Aharonov-Bohm oscillations has been observed before in
various materials, including graphene \cite{graphene_1, graphene_2,
temp_2}, and has been attributed to the loss of phase coherence due
to thermal averaging \cite{temp_2}. With increasing temperature, the
sample resistance decreases (Fig. 4(b)) but this change is not
significant ($< 10$\%), whereas the decrease in $R_{AB}$ is large
($> 200$\%). The diminishing amplitude of the oscillations $R_{AB}$
reflects the loss of phase coherence with increasing temperature.
We found all the features of the samples to be robust and the data
were reproducible even after several warm-up and cool-down cycles
over months.

Our work represents a clear demonstration of quantum interference between
electrons and holes and the efficient transmission of phase information
between the electron-type and hole-type regions. These measurements can also
give a direct estimate of the lower limit of the phase coherence length
at the Dirac point.

\end{document}